\documentstyle[12pt]{article}
\begin{document}
\pagestyle{headings}
\thispagestyle{empty}
\fussy
\parskip=0cm
\parindent=0cm

\def\ab{$\sim$}

\def\deg{$^{\circ}$\hspace*{-.1cm}}
\def\min{$'$\hspace*{-.1cm}}
\def\asec{$''$\hspace*{-.1cm}}
\def\hii{H\,{\sc ii}}
\def\hi{H\,{\sc i}}
\def\ha{H$\alpha$}
\def\hb{H$\beta$}
\def\hd{H$\delta$}
\def\hei{He\,{\sc i}}
\def\heii{He\,{\sc ii}}
\def\hg{H$\gamma$}
\def\sii{[S\,{\sc ii}]}
\def\siii{[S\,{\sc iii}]}
\def\oiii{[O\,{\sc iii}]}
\def\oii{[O\,{\sc ii}]}
\def\sm{$M_{\odot}$}
\def\slum{$L_{\odot}$}
\def\mdot{$\dot{M}$}
\def\x{$\times$}
\def\sec{s$^{-1}$}
\def\cm2{cm$^{-2}$}
\def\mcube{$^{-3}$}
\def\lam{$\lambda$}
\def\av{$A_{V}$}

\def\sk{Sk\,-71$^{\circ}51$}

\begin{center}

{\bf An Etymological Dictionary of Astronomy and Astrophysics \\

English-French-Persian}  \\

\vspace*{0.5cm}

M. Heydari-Malayeri\\

\vspace*{0.2cm}

LERMA, Observatoire de Paris, \\
61 Av. de l'Observatoire, \\
F-75014 Paris, France.

\end{center}

\vspace*{.5cm}

This online dictionary presents the definition of classical as well as advanced 
concepts of modern astronomy. Moreover, each English entry is accompanied by its 
French and Persian equivalents. The dictionary is intended to be helpful to 
professional as well as amateur astronomers. As a notable particularity, it also  
provides a detailed etymology of English and Persian terms. 
The etymological material contained in this work may interest linguists, 
in particular those concerned with the evolution of Indo-European languages, 
especially with that of their Iranian branch. \\

Apart from educational and outreach objectives in the field of astronomy, 
one of the main aims of this work is to contribute to the Persian language
by creating a comprehensive dictionary of astronomy and astrophysics.  Some of the basic 
premises, on which this project is based, are the following: \\

1. Persian, an Indo-European language with a rich literature and a
long written record, which goes back to about 1500 B.C. in its oldest
Avestan form, should be given support. In the present age of
exponential scientific/technological developments, the languages which
are incapable of expressing new concepts are unfortunately doomed to
disappear. It would be a dramatic loss if historical languages, which
have made important contributions to  human culture and
civilization, and therefore belong to common human heritage, died out. \\

2. The status of terminology in the field of astronomy and
astrophysics, and more generally in physics, is not satisfactory in
Persian. The author wishes to contribute to improving the situation,
and hopes that this initiative will prompt others to take up the task
of profoundly thinking about the causes, thus coming up with solutions
which will enrich the Persian astronomical terminology, and in a
broader extent the whole Persian language.  \\

3. Languages are vehicles of culture and diversity. Imagine a
magnificent garden with lots of different flowers, each with its
proper color, shape, and perfume; and compare it with another garden
filled uniquely with one kind of flower. No matter how lovely the
latter garden may be, it seems poor insofar as it dramatically lacks
diversity, and therefore constrains the freedom of choice. Similarly,
a night sky in which all the stars shine with the same brightness and
color is not only less romantic, it is also unfathomable, since the
differences necessary to study the nature of stars are lacking.  \\

4. In order to take up this linguistic challenge, the whole
capability of Persian should be used: not only its modern literary
heritage, but also the resources of Middle Persian (A.D. $c.$ -300 to +700), 
Old Persian (A.D. $c.$ -600 to -300), and Avestan
(A.D. $c.$ -2000 to -300). Further linguistic tools are needed to meet
the goal. The mine of various Persian dialects will be of great help.  
They have preserved very old Indo-European forms which are
missing in Modern literary Persian, and sometimes even possess terms
which are reminiscent of Proto-Indo-European roots whose Avestan and
Old Persian counterparts are extinct.  \\

5. It is only by employing all these means together in a general
scheme that Persian can own a powerful and efficient
scientific/technical language. Sticking merely to Modern Persian, as
traditionalists and conservatives do, seems highly inadequate and
therefore has little chance to succeed. Evidence of this is the
present dissatisfactory state of the Persian terminology in its
confrontation with an ever-increasing number of foreign terms and also
the lack of any workable project to solve the problems. In fact the
restriction to Modern Persian implies over-using a relatively limited
and incomplete word-base by multiplying combinations among a small
number of possibilities. That approach, which lacks solid linguistic
foundation and is rather ideologically motivated, can be likened to
tinkering ({\it bricolage}) when an overhaul is needed. It should also be
stressed that the most significant advances in Persian terminology in
recent years are essentially due to the reintroduction of forms and
affixes from ancient Iranian languages.  \\
 
6. Being Indo-European, Persian can luckily benefit from the model
as well as the past experience of the European languages which have
managed to produce their present powerful terminology system. Persian
has lots of cognates with the European languages and uses similar
word-forming patterns. As far as experience is concerned, after the
Renaissance, European intellectuals and scholars made use of the
reservoir of Greek and Latin in order to coin new concepts. Persian
can proceed in the same way with its ancestors, all the more so since
it has the advantage of profiting from recent linguistic findings. In
particular, Sanskrit, which is a sister/brother of Avestan/Old
Persian, can be of great help.  \\

The various characteristics of scientific language and terminology are discussed in the 
Introduction, where the criteria used in constructing Persian counterparts are 
extensively explained . \\

For detailed information and access to the dictionary, see:

\begin{center}  
{\bf 
http://aramis.obspm.fr/\ab heydari/dictionary/Intro.html
}
\end{center}

\end{document}